\documentclass[preprint]{revtex4}
\usepackage{graphicx}
\usepackage{epsfig}
\pdfoutput=1

\begin{document}
\title{Maturation of siRNA by strand separation:
Steered Molecular dynamics study}
\author{ Rakesh K Mishra$^{1}$, Sanchita Mukherjee$^{2}$, Dhananjay Bhattacharyya$^{*1}$ }
\affiliation{
$^{1}$ Computational Science Division, Saha Institute of 
Nuclear Physics Kolkata, 700064, India.\\
$^{2}$ Department of Biological Sciences and Centre for Climate and Environmental Studies, 
Indian Institute of Science, Education and Research-Kolkata, Mohanpur 741246, India }

\begin{abstract}
 RNA interference, particularly siRNA induced gene silencing is becoming 
an important avenue of modern therapeutics. The siRNA is delivered to the 
cells as short double helical RNA which becomes single stranded for forming the 
RISC complex.
 Significant experimental evidence is available for 
most of the steps except the process of the separation 
of the two strands.  We have attempted to understand the pathway for 
double stranded siRNA (dsRNA) to single stranded (ssRNA) molecules using 
steered molecular dynamics simulations. As the process is completely 
unexplored we have applied force from all possible directions restraining 
all possible residues to convert dsRNA to ssRNA. We found pulling one 
strand along helical axis direction restraining far end of the other 
strand demands excessive force for ssRNA formation. 
 Pulling a central 
residue of one strand, in a direction perpendicular to the helix axis, 
while keeping the base paired residue fixed requires intermediate force 
for strand separation. Moreover we found that in this process the force 
requirement is quite high for the first bubble formation (nucleation energy) 
and the bubble propagation energies are quite small. We hypothesize this could 
be the mode of action adopted by the proteins in the cells. 
\end{abstract}

\maketitle

\section{introduction}
Ribonucleic acid (RNA) interference (RNAi) was first time discovered 
in plants, but it was not noted in animals until Fire and Mello 
demonstrated that double-stranded RNA (dsRNA) can cause greater 
suppression of gene expression than single-stranded RNA (ssRNA) in 
Caenorhabditis elegans \cite{Fire}. Due to the excellent gene silencing 
potential of RNAi, it has attracted broad attention to 
exploit its capabilities. In recent years, RNAi has become more 
and more important in gene silencing and drug development because of 
its high specificity, significant effect, minor side effects and ease 
of synthesis \cite{Dogini}. When dsRNA enters the cell, it is first 
cleaved into short double stranded fragments of $20-23$ nucleotide 
silencing RNAs. These cleaved products have been recognized as 
the small interfering RNAs (siRNAs) in the form of double stranded helices. They 
are generally named as passenger strand and guide strand. Presumably the double helix
also remains complexed with 
RNA-Induced Silencing Complex (RISC). In the RISC, the guide strand 
of siRNA pairs with a complementary sequence in a messenger RNA (mRNA) 
molecule and induces cleavage of mRNA by enzyme Argonaute. Thus, the 
process of mRNA translation can be interrupted by siRNA \cite{Sashi, Martin, Meist, Li}. 
Since rational design of siRNA can specifically inhibit endogenous and 
heterologous genes, it can modulate any disease-related gene 
expression. Following this strategical revelation, several synthetic 
siRNA are being designed with desirable sequences to inhibit 
any target gene expression \cite{Lopez,David,Fell}. The siRNAs undergoe 
further processing inside the cell, where, one strand or part of one 
(say guided strand ) gets separated from the other strand (say passenger strand). 
This separation takes place by the force generated by some enzymatic 
reaction. The naturally coded microRNA (miRNA) also goes through similar 
steps for their action. Thus, the structure and force involved in the 
separation of the strands become one of the important aspects 
to deal with the efficiency of siRNA.

Double-helical deoxyribonucleic acid (dsDNA) has been widely studied 
with respect to strand separation based on experiments and theory \cite{Garima,Kumar,
Essvaz,Bockel,Strunz,Schum,Danilo,Hatch,Cdan,Kuher,Marko}, where 
researchers have studied the effects of mechanical force on the 
structural changes of dsDNA. It has been shown that differences 
in the chemical structure of dsDNA and dsRNA molecules affects the 
intra-strand distances \cite{Alberto}. Recardo {\it et al.} \cite{Recardo} 
have studied the effects of force using optical and magnetic 
tweezers on the stretching of dsRNA and compared it with dsDNA 
results. Lipfert {\it et al.} \cite{Lipfert} have studied the 
effects of force and torque on the structural changes of long 
stretch of dsRNA and pointed out striking differences between dsRNA 
and dsDNA. Unfolding of compact structure of RNA was also studied 
recently by various groups using experiment and simulation \cite{Mandal,Vishnu, Savinov, Manju}.

Nevertheless, siRNA evolves as one of the unprecedented small 
bio-molecule, unlike large polymeric DNA or various RNA motifs, 
requiring broad study of structural changes under the application 
of external mechanical force. Oligomeric siRNA, having 20 to 22 
base-pairs and UU overhang in both the strands requires special 
attention in its structural changes during unzipping. It may be 
mentioned that natural microRNAs (miRNA) also require unzipping 
after they are processed by DICER protein, which may require some 
assistance from proteins. We approach this novel problem of siRNA 
strand separation by focusing our study on the opening of both the 
strand of siRNA under the application of external mechanical force. 
This mechanical force might come from some protein, such as Argonaute, DICER  
or some other RNA binding protein, within the cell. However, we could 
not find any report on action of such protein in the literature. 
It may be noted that partial strand separation of dsDNA also takes place during 
transcription initiation. It was found that the sigma-factor of RNA polymerase
is responsible for that \cite{Saecker, Zuo, Aayatti}. Thus, separation of strands of
 double helical nucleic acid chains is an important aspect for understanding
several biochemical pathways.
In this report, 
we present the effects of the pulling with the constant velocity 
steered molecular dynamics (SMD) simulations under different 
protocols as shown in Figure 1, broadly classifying them as axial 
pulling and unzipping. We provided a comparative study using all 
atom MD simulation, which may help in characterizing the structural 
changes that occur during the pulling of double-stranded siRNA. We 
address the problem with extensive studies of structural parameters, 
hydrogen bonds (H-bonds) disruption and stacking interactions. 
\begin{figure}[t]
\includegraphics[height=3.2in, width=3.3in]{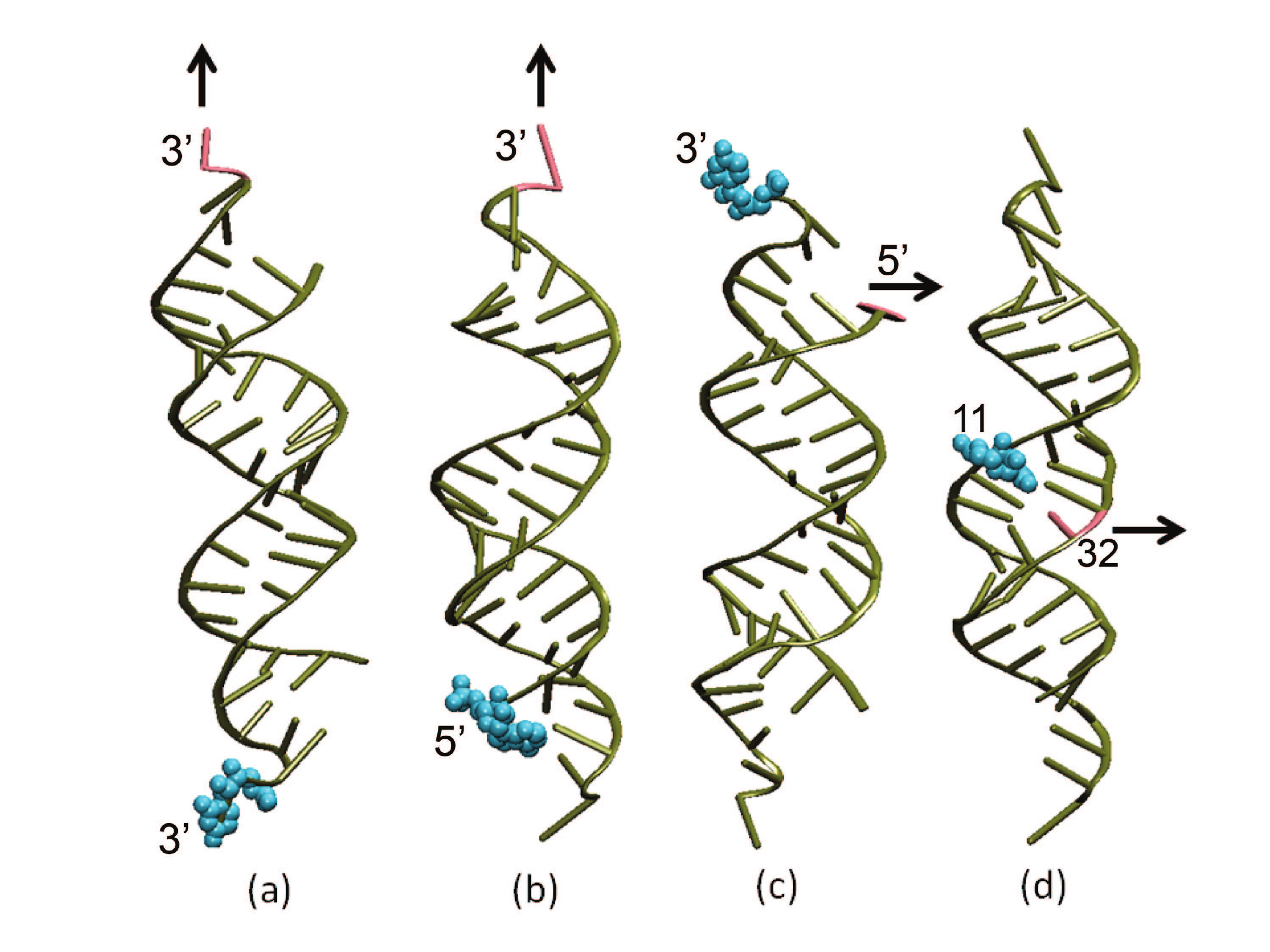} \\
\caption{Schematic diagram showing different protocols studied in this 
report: (a) Shows the Axial Rupture system, where force is applied along the 
helical direction. In this case force is applied at $3'$-end of one strand 
keeping $3'$-end of it's complementary strand fixed. (b) Shows the Axial 
Stretch model, where force is applied along the axial direction of the 
system. But, in this case pulling is applied at $3'$-end of one strand 
while $5'$-end of the same strand being fixed. (c) Shows Terminal Unzip model, 
where force is applied perpendicular to the helix direction. In this case, 
pulling is applied at $5'$-end of one strand keeping $3'$-end of its 
complementary strand fixed. Here, both pulling and fixed ends are in the 
same side of the duplex. (d) Corresponds to the Central Unzip model, where 
force is applied perpendicular to the helical direction at the central 
residue of the system.}
\label{fig-1}
\end{figure}
The present studies, which are involved in calculation of path dependent 
force and disruption of H-bonds by various protocols may shed the light to 
provide future perspective of binding proteins with siRNA for gene silencing.  
The analyses may also be important for designing more effective siRNA sequences.

The paper is organized as follows: in Section II, we describe the computational 
protocol for simulation and analysis. Section III contains analysis of the 
measured force on the strands and the reason behind such forces. In Section IV,
we have discussed variation of different structural parameters of the RNA. 
Section V describes implication of the studies and how our results are 
consistent with the available experimental data.

\section{Model and Method}
\subsection{Equilibrium MD simulation:}
{\bf Starting system :} 
Well studied siRNA crystal structure with PDB ID 2F8S \cite{Yuan} is taken for all the 
simulations. The structure is comprised of 22 nucleotides on each strand of the 
duplex with characteristic UU overhang in both the $3'$-ends. The self-complementary 
sequence is, $ 5' (AGACAGCAUAUAUGCUGUCUUU)_{2} $. The dimension of the duplex is 
$\approx$ $8.0$ nm in length in normal double helical form. 

{\bf Protocol:}
All molecular dynamics simulations were carried out using GROMACS 5.1 package 
\cite{Abraham} and CHARMM36 Force Field \cite{Best}. The complete siRNA double 
helical structure, without the Argonaute protein interacting through one end of
 the RNA, was considered including the $5'$-terminal phosphate groups. The siRNA is solvated with 
TIP3P water model in cubic box with sufficient dimension (10 nm $\times$ 10 nm $\times$ 10 nm) 
and neutralized with 44 Na$^{+}$ ions. The system is then subjected to energy 
minimization by steepest descent method to eliminate initial stress. For 
initial equilibration, standard protocol of 100 ps each of NVT and NPT simulations 
were done \cite{Abhijit}. Position restraints was applied to the RNA atoms 
during equilibration of the system for both NVT and NPT processes. The siRNA 
and non siRNA atoms were coupled to separate temperature coupling baths, 
maintaining 300 K using Berendsen weak coupling method \cite {Beren}. 
Final 100 ns NPT production MD run was conducted in absence of any restraints. 
In the production run, the Nos´e-Hoover thermostat \cite{Nose,Hoover}  was used 
to maintain temperature and the Parrinello-Rahman barostat \cite{Nose, Hoover, Nose2,Rahman} 
was used to isotropically regulate pressure. Periodic boundary conditions 
(PBC) were employed for all simulations and the particle mesh Ewald (PME) 
method \cite{Darden} was used for long-range electrostatic interactions. 
The simulation time step was set to 2 fs with LINKS algorithm to maintain 
bond lengths involving hydrogen atoms. Final structure from the end of 100 ns 
equilibrium trajectory was used as starting configurations for pulling simulations.

{\bf Steered MD simulation:}

Four different steered molecular dynamics simulations (SMD) were carried out with 
equilibrated conformations of siRNA along with solvents at 300 K. Henceforth we would 
refer to them as: (i) Axial Rupture, (ii) Axial Stretch, (iii) Terminal 
Unzip and (iv) Central Unzip. In case of Axial Rupture, shown in Figure 1a, force 
is applied at the $3'$-terminal residue of one chain keeping the $3'$-terminal residue 
of it’s complementary strand fixed (immobile) to their original position. For Axial 
Stretch (Figure 1b),  we have fixed the $5'$-terminal residue of one strand and 
applied force on the $3'$-terminal residue of the same strand along the 
helical direction of the system. In case of Terminal Unzip, shown in Figure 1c, we 
have fixed the $3'$-terminal residue  of one strand and force is applied on the $5'$-end 
of its complementary strand along a direction perpendicular to the helical axis. 
Here both the $3'$- and $5'$-terminal residues are in the same side of the double 
helix. In Central Unzip case (Figure 1d) the force has been applied in the central point 
(residue 11) of one strand keeping the central residue of the complementary strand 
fixed. This is also the case, where force is applied perpendicular to the helical 
direction. In case of pulling simulations, big rectangular boxes with dimensions
sufficient to satisfy minimum image convention for complete separation of
siRNA were generated. This provided space for the nearly elongated single stranded RNA along the Z-axis for 
the rupture and along Y-axis (perpendicular to the Z-axis) for the unzipping.
We have adopted the boxes of the size 15 nm $\times$ 15 nm $\times$ 54 nm,  
15 nm$\times$ 15 nm$ \times$ 54 nm,  15 nm $\times$ 40 nm $\times$ 15 nm  and 
15 nm$\times$ 40 nm $\times$ 15 nm, for Axial Rupture, Axial Stretch, Terminal Unzip 
and Central Unzip respectively.  These boxes were filled with TIP3P model of
explicit water with adequate Na+ counterions to neutralize the systems.  
Equilibration was performed for 200 ps NVT and 10 ns NPT simulations, 
using the same methodology described above prior to SMD simulations.

{\bf Protocol:}

SMD is based on applying external forces to particles in a selected direction 
by adding a spring-like restraint, thus imitating directly the basic idea of 
an AFM experiment through optical or magnetic tweezers. 
The SMD simulations with constant velocity (CV) stretching (SMD-CV protocol) 
were carried out by fixing one of the residues and applying external forces 
to the dummy atoms attached to center of mass of another residue (SMD residue) 
with a virtual spring. After several test simulations, we adopted a spring constant value of 
$1000 \times {kJ} \times {{mol^{-1}} \times {nm^{-2}}}$ 
and a pulling rate of $ 0.0008 \times nm \times {ps^{-1}} $. 
The force experienced by the pulled terminal residue, $ F $ is defined 
as $F(t) = k(vt - x)$ where, $ x $ is the displacement of the pulled atom 
from its original position, $ v $ is the pulling velocity, and $ k $ is 
the spring constant. 
The direction of pulling was such 
that the end-to-end distance always increased, {\it i.e.}, the SMD residue 
was pulled away from the fixed residues.

{\bf Analysis:}

Analysis of the trajectories, including finding number of H-bonds were done
by GROMACS 5.1 \cite{Abraham}. Base-pair orientation parameters and stacking geometry
were analysed by NUPARM \cite{Bansal,Shayantani,Pingali}.

\begin{figure}[t]
\includegraphics[height=3.5in, width=5.0in]{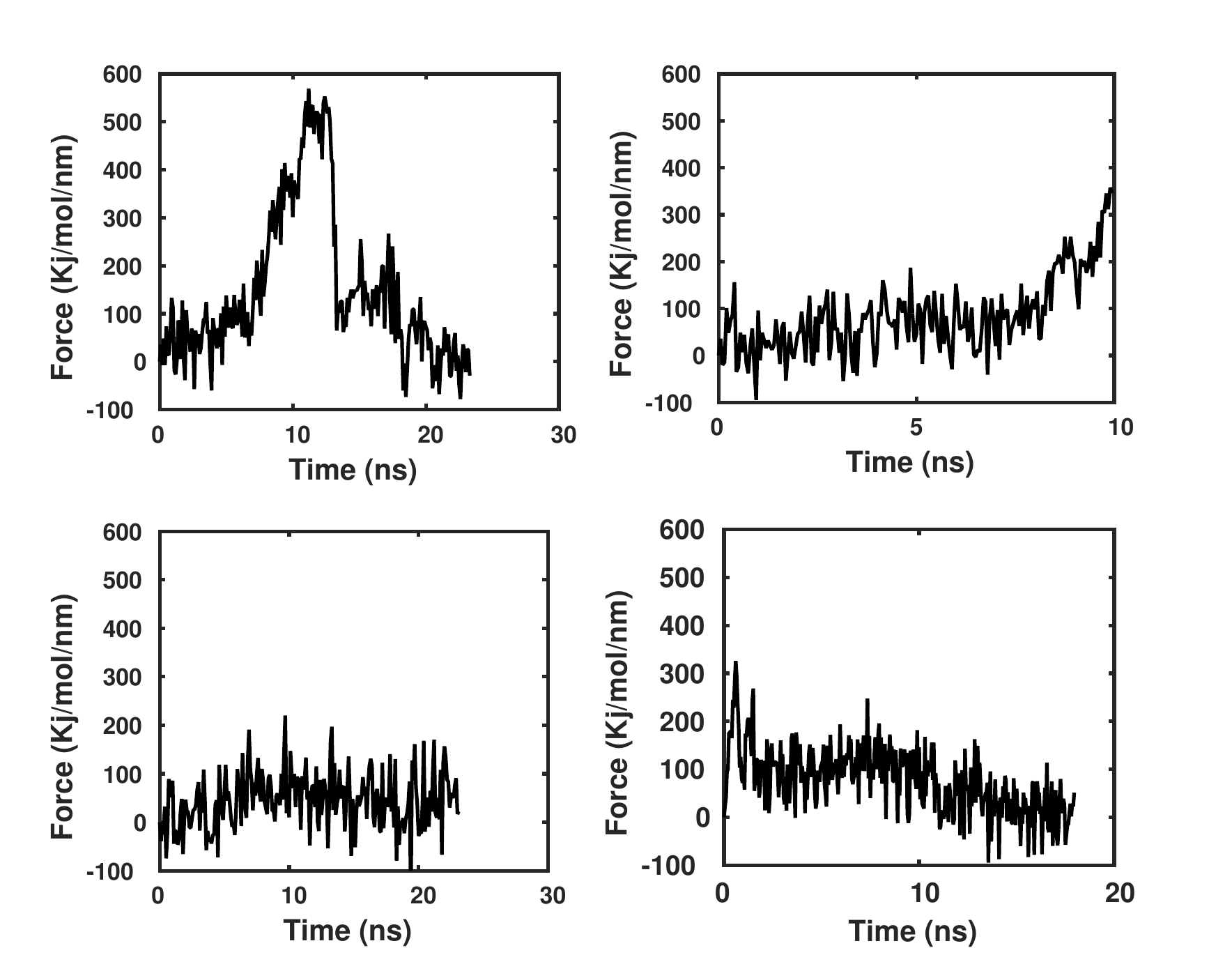}  
\caption{Shows the variation of force experienced by the system with the 
time under different protocol as shown in Figure 1. (a) Corresponds to the 
variation of force with time for the Axial Rupture model. (b) Corresponds the 
variation of force for the pulling of the Axial Stretch model. (c) Corresponds 
the variation force for the pulling of the Terminal Unzip model and (d) 
Corresponds the variation force applied for the Central Unzip model of siRNA.}
\label{fig-2}
\end{figure}
\section{Results analysis}
We first look at the stability of the equilibrium MD simulation. Previous 
equilibrium molecular dynamics studies by several groups had observed that separation of the strands of 
siRNA took place during interaction with graphene or carbon 
nanotube \cite{Jung, Prabal,Santosh, Landry, Ghosh}. It was also demonstrated 
that such separation of strands did not happen in case of double stranded DNA\cite{Prabal, Santosh}, thus attributed this siRNA separation to specific interaction between
carbon nanomaterial and siRNA. Our recent study, however indicate siRNA remains in double 
helical form in physiological environment \cite{Abhijit}. This indicates that 
external force is possibly needed to compel the separation of strands in siRNA. 
The results compliments our equilibrium simulations, which also demonstrates 
only moderate RNA breathing in equilibrium MD throughout the $100$ ns production run. 
We have taken the final equilibrated structure of siRNA duplex from the equilibrium  
MD simulation for further force induced SMD simulations. In SMD or center of 
mass (COM) pulling the system is biased to demonstrate the behavior toward a 
particular phenomenon. Application of an external force to cause displacement 
in the simulated system allows for the calculation of work, a path-dependent 
quantity. For opening of the strands of the siRNA we adopted two standard 
protocols as described in method section (Figure 1), one is to apply force along 
the helical direction of the system and the other is to apply the force 
perpendicular to the helical direction of the system, the later can be generalised 
as unzipping.

We have measured the forces experienced by the siRNA, which vary with time 
for the different model systems (Figure 2).  Figure 2a shows the variation of 
force with time for the Axial Rupture model. We observe nearly linear variation
of extension with simulation time (Figure SI 1), hence the force {\it vs}. extension
curves also look very similar.  In the case of Axial Rupture one strand is pulled 
along helical axis direction keeping the far end of the other strand constrained. 
During the initial phase of pulling, the $3'$-terminal single stranded UU residues 
adopt stretched out conformation, which does not require any extra effort.  
The double helical structure is also not affected by such conformational change
of the single stranded region.
After this phase, varying peaks of the curve reveal  that structural transition  
occurs and the system starts to change from its double-helical form. One expects 
the hydrogen bonds (H-bonds) between the complementary bases to break at this 
moment. The stacking interactions between successive base-pairs also may get 
affected during this second phase of simulation with large force. The third 
option is that both take place simultaneously during this phase.  In order 
to understand the mechanism, we have analysed number of H-bonds present in the 
system at each time frame (Figure 3). Total number of H-bonds continue to 
decrease with time in the first phase, {\it i.e.} upto 10 ns (Figure 3a, blue 
curve). After about 10 ns, the number of H-bonds do not reduce significantly 
with time, possibly indicating most of the phase transitions took place by 10 ns. 
However, the number of H-bonds in the double helix does not reduce to zero value 
within this time. As seen in Figure 2a, the force increases at this point of 
time when number of H-bonds between the two strands appear to increase slightly. 
After this phase transition, number of H-bonds slowly reduces to zero when 
the strands dissociate completely (Figure 3a, blue line).  But, after this 
critical interaction the system breaks and the requirement of an additional 
force starts to decrease. In this time duration, force 
reduction from maximum to minimum reveals the structural change of the system 
from the bound double helical state to the completely unbound ssRNA state. Both the strands 
become almost separated, where most of the base-pairs are broken. It may be 
noted that the unfolded single stranded chains has capability to form intra 
strand H-bonds and hence, the blue line does not reach to zero value. This can 
be visualised by the various snapshots shown in the first row of Figure 4.

Variation of force  with time for the Axial Stretch model system is shown in 
Figure 2b, where the constrained point and pulling end are on the same strand 
and pulling is done along the helical direction (Figure 1b) somewhat similar 
to the Axial Rupture model. Variation of force with time, however, is found to differ significantly
from that of Axial Rupture model. Here, slight increase in the force during  
the last phase of SMD simulation is presumably due to conformational transition 
of the pulled strand to somewhat all-trans geometry, which is obviously not 
energetically favorable, especially for nucleic acids \cite{Bourne}. 
\begin{figure}[t]
\includegraphics[height=2.0in, width=5.0in]{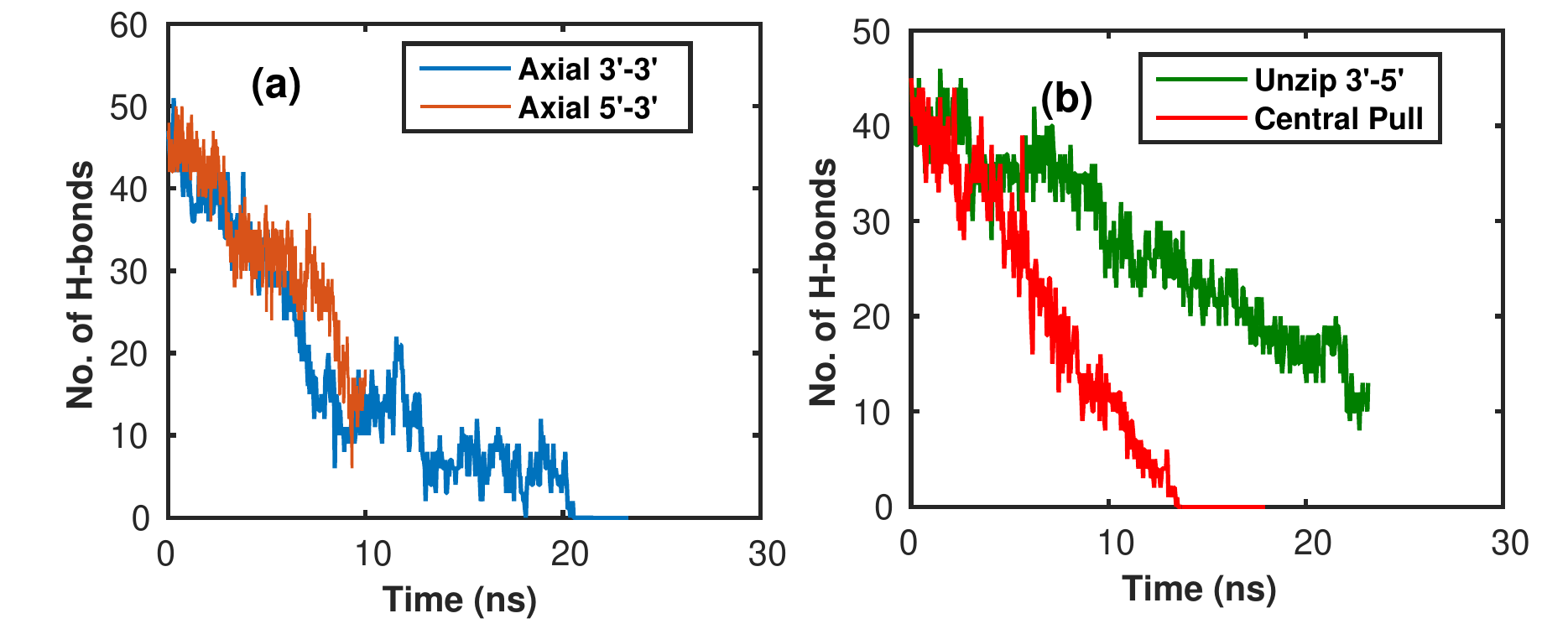} 
\caption{ Variation of number of H-bonds in between the two strands of siRNA 
with time during (a)  Axial Rupture (blue line) and Axial Stretch (yellow 
line) SMD simulations and (b) Terminal Unzip (green line) and Central 
Unzip (red line) SMD simulations.}
\label{fig-3}
\end{figure}
Nevertheless, in this case completely separated strands do not arise 
because pulling is done on the strand which is also fixed at the other 
end. But, it is noticeable that, even then force induced breakage of 
most of the base-pair H-bonds between the complementary strands take 
place (yellow line in Figure 3a). This can also visualised by the 
snapshots shown in second  row of Figure 4. The helical structure is 
converted to ladder like form (Figure 4), but most of the bases remain close 
to their complementary ones of the opposite strand. Thus, some H-bonds 
were retained till the end of the SMD simulation and further pulling 
was impossible as that would need to break or stretch the covalent bonds. Possible such effort took place after around $8$ ns, showing increase in the measured force.

Variation of unzipping force with time for the Terminal Unzip model system
is shown in Figure 2c. Here, unzipping starts from one end and propagates to 
the far side progressively. This kind of strand separation was also observed 
by several groups in MD simulations of dsDNA at elevated temperature 
\cite{Wong, Zgarbova,Sangeeta} or partially even at physiological condition. This can 
be termed as extension of fraying or peeling effect. In this case 
intact base-pairs are breaking gradually and progressively one by one, which 
causes the system to experience almost equal small force in steps until 
complete separation of the strand. And hence, variation of force is 
nearly at constant small value at all time steps. The progression of simulation can be seen 
by the different snapshots shown in first row of Figure 5. The base-pairs break 
continuously as time progresses, which is also reflected in number of H-bonds 
(Figure 3b, green curve). Significant reduction of number of H-bonds was 
not observed till 8 ns, as the single stranded UU residues were changing 
to stretched out conformation during this phase.

We observe unique result on the Central Unzip model system (Figure 1d), 
where force is also applied perpendicular to the helical direction of the RNA 
but in this case at the central residue (11th residue) of one of the strands 
keeping the paired residue of the complementary strand immobile. Here, also 
the variation of force experienced by the system with time is qualitatively 
similar to that of terminal-unzip model system. However, it is notable that the measured force is significantly larger 
(nearly 300  {kJ} {{mol$^{-1}$} {nm$^{-1}$}}) in the initial phase (Figure 2d) as compared to the other systems.
 After the initial phase 
the force reduces to smaller magnitude around  100  {kJ} {{mol$^{-1}$} {nm$^{-1}$}} 
and even smaller values. This initial increase of force can be explained as 
initiation of base-pair opening and the later as propagation of base pair 
opening to both sides of the central one.  This is equivalent to nucleation 
energy for cooperative transition from helix to coil state. The nucleation 
energy is quite high, as it would disrupt a base pair (at least two hydrogen 
bonds) and two stacking interactions between the pulled base pair and its two 
neighboring base-pairs on both sides. The second type of force is supposedly 
stronger than the base pairing energy and it is doubled also \cite{Peter}. 
Once the H-bonds of the central base-pair break and the stacking between the 
central base-pair and its neighbouring base-pair are disrupted, the neighbouring 
base-pairs can have fraying like effect. In other word these neighbouring base-pairs 
come to contact with solvent water. Hence, These bases can form H-bond with the 
complementary bases or with solvent water molecules in a competitive manner. The trajectory of this 
model system can be visualised by the snapshots shown in the second row of Figure 5. 
Furthermore, the propagation stage is quite faster as compared to the other SMD 
results, as two base-pairs break together, {\it i.e.}, C+1 and C-1 after base 
pair C breaks (where C is the central base-pair) then C+2 and C-2 break and 
so on. Hence, they can now easily become single stranded breaking the Watson-
Crick base pairing  and stacking interactions, requiring small amount of force.
\begin{figure}[htp]
\includegraphics[height=3.6in, width=6.7in]{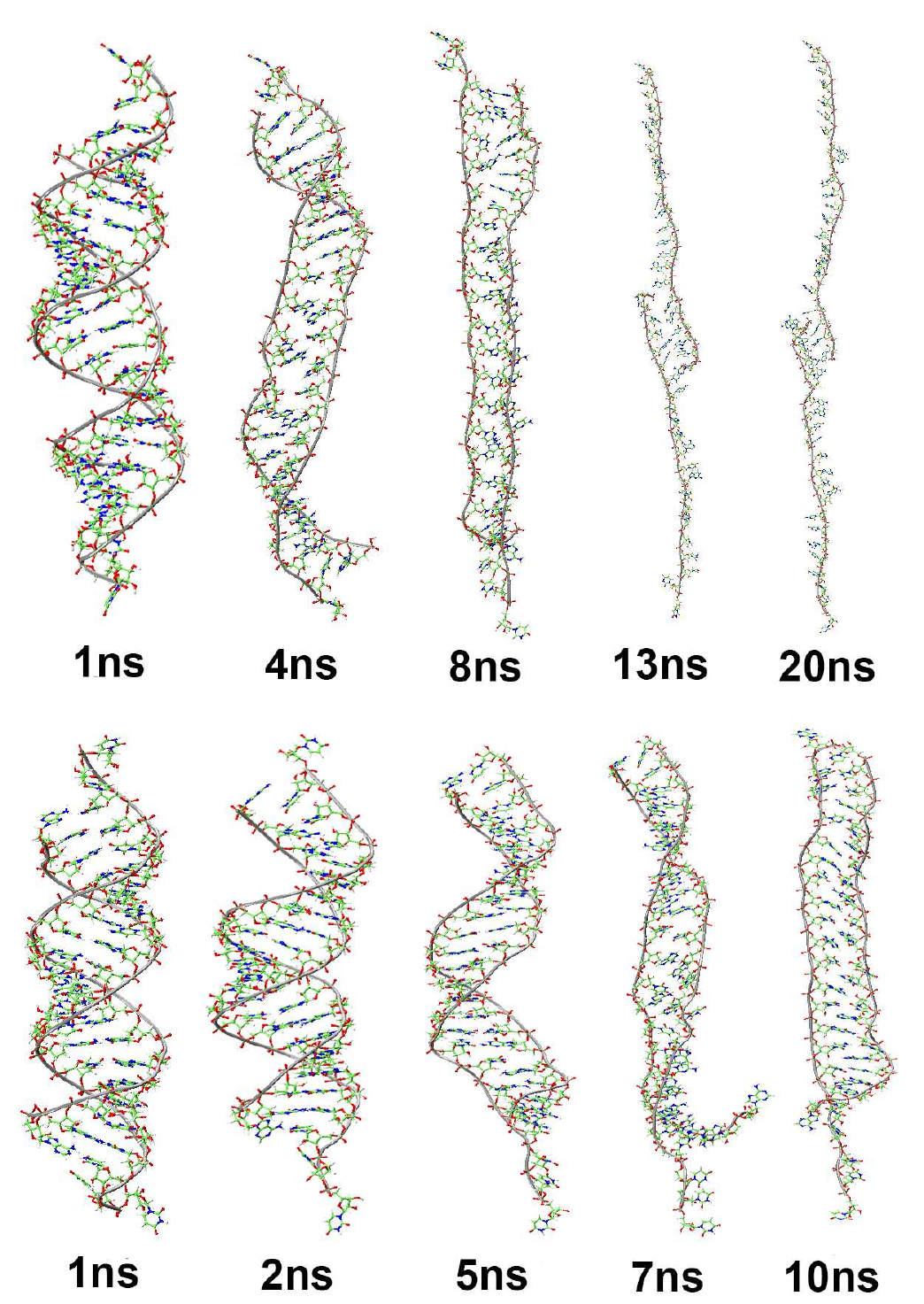} 
\caption{ Panel of axial force application: First row corresponds to the snapshots of
structural transition for Axial Rupture model system. Second row corresponds to the
snapshots of structural transition for the Axial Stretch model system. }
\label{fig-4}
\includegraphics[height=3.6in, width=6.7in]{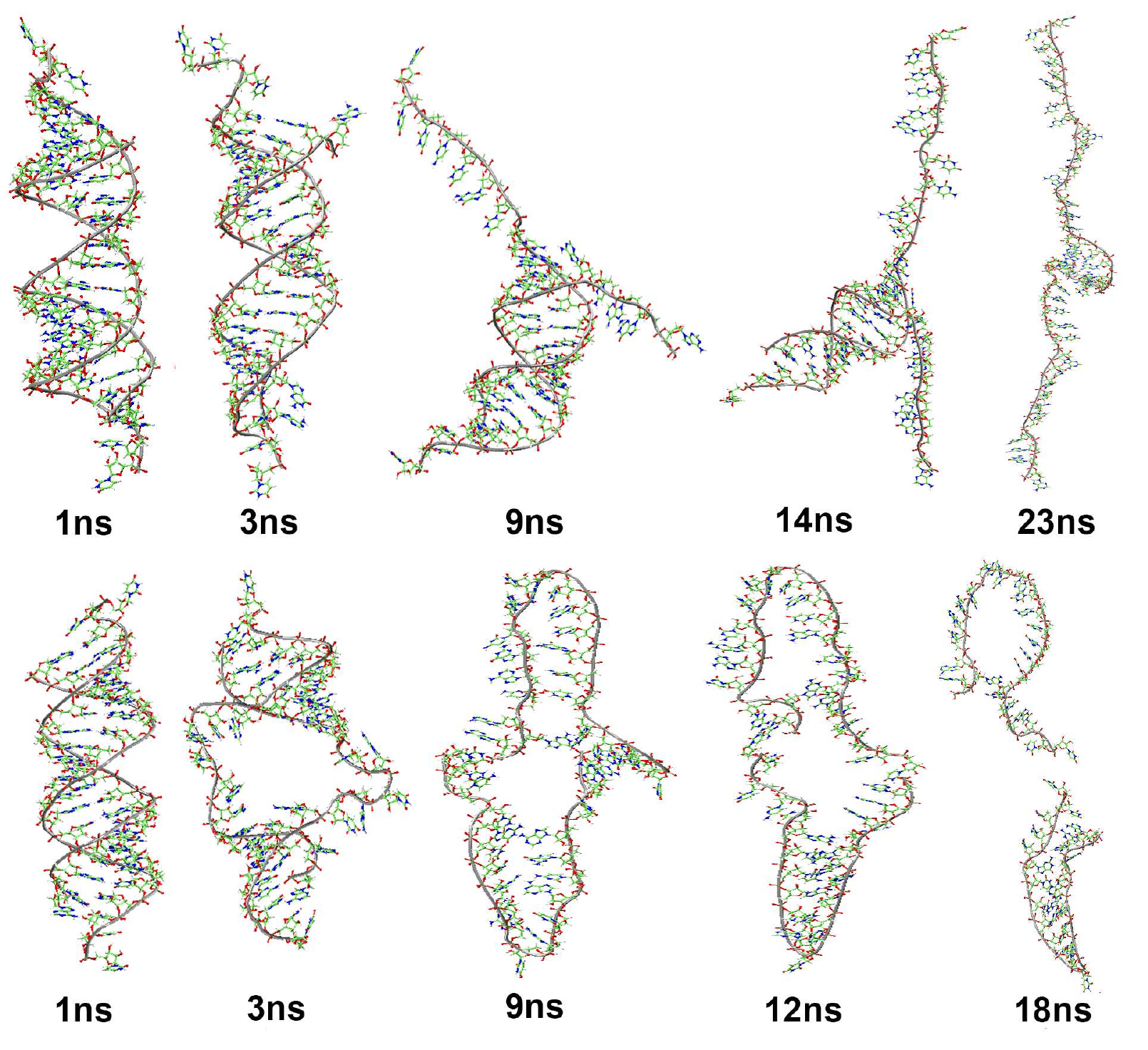} 
\caption{ Panel of force applied to the unzipping model systems: First row
corresponds the snapshots of structural transition for the Terminal Unzip model
system. Second row corresponds the snapshots of structural transition for
the Central Unzip model system. }
\label{fig-5}
\end{figure}
\section {structural transition}
As indicated above, the H-bonds between the complementary bases in a base-pairs 
breaks during force induced SMD simulations. Thus the bases do not remain coplanar 
to each other and the other degrees of freedom of the bases also increase  beyond 
their regular values. Quantitative analysis of these degrees of freedom of the bases 
with respect to paired ones can be done by the six IUPAC-IUB recommended intra 
base-pair parameters \cite{Olson}. We have therefore looked at shear, open 
angle and stretch values, which are related to the H-bonding features. Similarly 
relative orientations of a base-pairs with respect to their neighbouring stacked 
ones also change significantly when the stacking interactions are disrupted. 
These can be analysed by tilt, roll, etc., inter base-pair local parameters. 
Effect of all these inter base-pair parameters can also be analysed by a 
composite parameter, namely stacking overlap, and we have analysed that also. 
 Variation of shear, base-pair overlap and twist, as representative 
parameters, are shown in Figure 6 and Figure 7 to compare all the systems. 
\begin{figure}[htp]
\includegraphics[height=4.0in, width=7.2in]{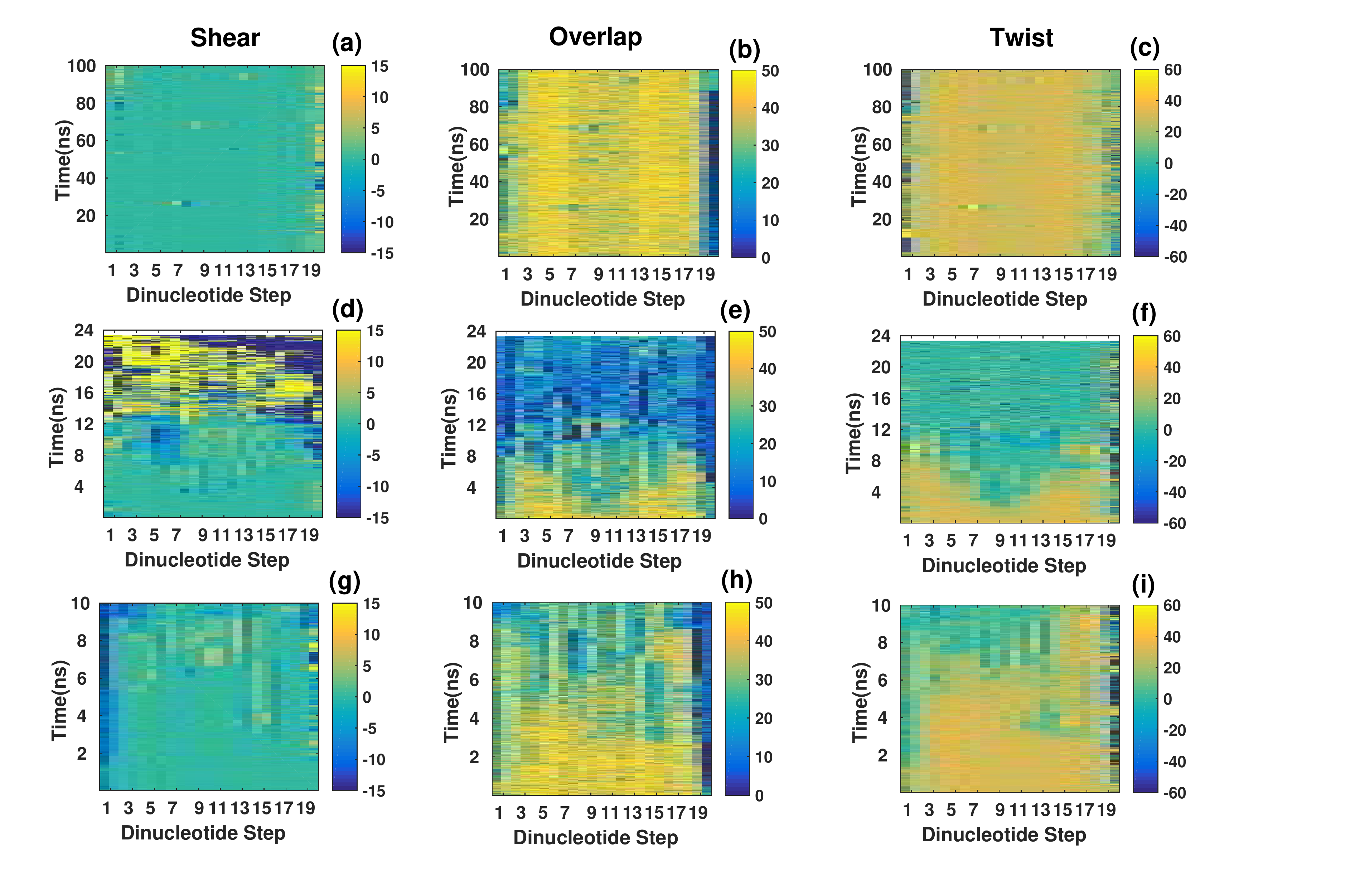} 
\caption{Variation of base pair parameters during equilibrium and axial pulling 
simulations. Residue numbers are shown in x-axis, time in y-axis and the 
graded color for values of (a) shear ({\r{A}}), (b) overlap ({\r{A}}$^{2}$)  
and (c) twist ($^\circ$), for equilibrium simulation. The shear, overlap and 
twist for Axial Rupture model are shown in (d), (e) $\&$ (f) while (g), (h) 
$\&$ (i) are the same for Axial Stretch.}
\label{fig-6}
\includegraphics[height=3.2in, width=7.2in]{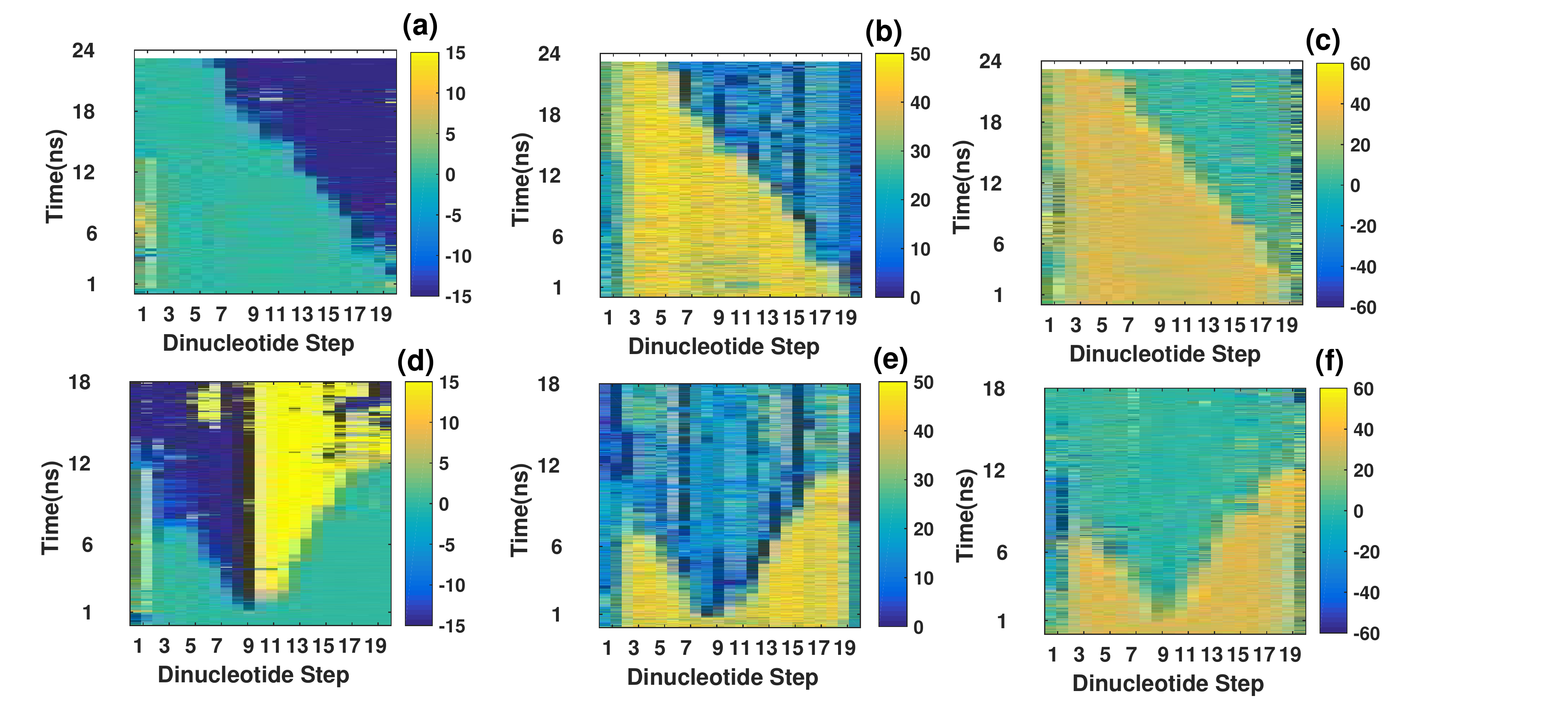} 
\caption{Base pair parameters for unzipping SMD simulations. The base-pair 
numbers are shown in x-axis, time (ns) in y-axis and the graded color represents 
values of: (a) shear  ({\r{A}}), (b) overlap ({\r{A}}$^{2}$) and (c) twist ($^\circ$) 
for the for Terminal Unzip model  and (c) shear ({\r{A}}), (d) overlap 
({\r{A}}$^{2}$) and (e) twist ($^\circ$) for Central Unzip 
model.}
\label{fig-7}
\end{figure}

The shear parameter gives information about relative movement of the bases with 
respect to the paired ones, indicating disruption of hydrogen bonds in a base 
pair, overlap provides information about stacking between two base-pairs and 
twist indicates ladder like structure formation, and hence is often related to 
stretching \cite{Alberto}. These are illustrated as three dimensional plots. 
It may be noted that shear values of good Watson-Crick base-pairs are around zero \cite{Shayantani}, 
twist value of A-RNA stretches are around 33$^\circ$ \cite{Bourne} and stacking overlap between 
successive base-pairs in RNA double helices are around 45 to 50 {\r{A}}$^{2}$ \cite{Pingali}
depending on base sequence. We have also analyzed variation of these parameters
during equilibrium MD simulations of the siRNA to understand the force induced
effects on the double helical structure (Figure 6a, b $\&$ c). 
From the 
figures, it is clear that the structure maintains almost ideal values of these 
parameter throughout the simulation, indicating that no separation of the 
strands took place in absence of any force or molecules like graphene or 
carbon nanotube. Furthermore, terminal fraying is also found to be minimum as
compared to other simulations using CHARMM force-field \cite{Sangeeta}, possibly
due to somewhat capping effect by the two single stranded residues at the two ends
of the double helix.

The changes in the parameters for the Axial Rupture system are shown in Figure 6d, e $\&$ f. 
Figure 6d illustrates that initially upto 4 ns shear values of all the 20 
base-pairs are near $0$ (sky blue color), indicating no disruption of any 
base-pair. After that minor disruption of base pairing can be seen with minor 
fluctuation of shear values until 10 ns. Here, the shear values of 5th, 6th 
and 7th base-pairs become more negative and can be seen to be more affected 
by the force. Base-pairs of $5'$-terminal residue adopt large positive shear 
and all the base-pairing near $3'$-terminal assume large negative shear. All 
the shear values start changing because of structural transition after 12 ns 
and afterwards huge fluctuation of shear (Yellow color corresponding to values 
around $10 - 15$  {\r{A}} or deep blue color for shear values around $- 15$
{\r{A}} ) indicate disruption of initial base pairing. Comparing with disruption 
of H-bonds (Figure 3a for Axial Rupture model) it can be concluded that, after 
breaking half of the total possible H-bonds (from $\approx$ 48 to 25) the 
disruption is steep linear after around 12 ns. Variation of overlap parameters 
(Figure 6e) illustrates continuous disruption of stacking for all the base-pair 
steps within 8 ns, and it is seen to initiate after 2 ns at the central region. 
However, even at this time (8 ns) most of the bases are paired to their 
complementary ones, as reflected from the analysis of shear. Thus, base-pairs opening and stacking disruption appear 
to be independent and unrelated events. For few instances, we found that, 
after complete breakdown of the stacking, again overlap value increases 
due to single stranded helix like structure formation. Such single stranded 
structure formation leads to stacking between successive bases, instead of 
stacking between successive base-pairs in double-stranded RNA. 
From Figure 6f it appears
that the twist values of the base-pairs at different helical positions start 
to fluctuate at diverse time points. The middle base pair steps (9th and 10th) 
faces twist disruption at as early as 2 ns. Comparing with the two other  
parameters, shear and overlap, we found twist to get disrupted earliest and 
become most sensitive to the applied force. It is noted that the central base-pairs 
are getting effected earlier possibly due to weaker base-pairing in the central 
region (AU rich sequence). These structural transitions can be seen with snapshots throughout the 
simulation in first row of Figure 4.

Structural parameter variations for the Axial Stretch model of same strand 
pulling are presented by Figure 6g, h $\&$ i. Interestingly, it can be 
observed that in this case values of the parameters have complete different 
signature of variation as compared to the Axial Rupture model. We found that 
the regular shear variation is maintained for much longer time of pulling 
for all the base-pairs as compared to Axial Rupture model (Figure 6d). 
However, somewhat larger fluctuation of shear values is observed for few 
base-pairs (14th to 17th) after 4 ns.  These base-pairs are also seen to be 
unstacked with respect to their neighbours from 4 ns (Overlap values reduce 
to around 20 {\r{A}}$^{2}$). As expected twist value of these base-pairs 
also reduce at the same time. Disruption in stacking overlap is also found 
less for Axial Stretch model system (Figure 6h) as compared to Axial Rupture 
model system (Figure 6e). Even at the end of Axial Stretch model simulation 
significant stacking overlap around 30 {\r{A}}$^{2}$ is found between successive 
base-pairs, indicating separation of the strands did not take place. 
Comparing with the breaking of H-bonds for this case (Figure 3a, yellow line) 
showing that even after the system is melted the number of H-bonds are still significant 
($ > 10$ number), which can be observed by different snapshots shown in second 
row of Figure 5. We find twist values (Figure 6i) are sensitive to the force 
as compared to shear and overlap in Axial Stretch model.  However, the 
fluctuation starts late as compared to Axial Rupture model. These differences 
in variations can be related to lesser disruption of H-bonds in Figure 3a 
for axial-stretch model, indicating the situation that complete force induced 
rupture is not taking place.

Variations of base pair parameter for Terminal Unzip model system are shown in 
Figure 7a, b $\&$ c. In this case the fixed end and the end at which pulling force 
is applied are on the same terminal side of the duplex and this unzipping mode 
is prominent in the signature of parameter variations.  Variation of shear 
(Figure 7a) indicates that the disruptions of base-pairs are completely 
dependent on the base positions. The terminal to the pulling end is first 
disrupted by the unzipping force, which starts to perturb the system from the 
pulling end gradually (19th base pair). Similar gradual disruption of overlap 
values can be observed from Figure 7b. This illustrates that the unzipping 
effect are progressive in nature and minimal effect is transmitted to the 
bases far from the current unzipped base pair. This is almost equivalent to 
extension of fraying effect seen earlier \cite{Sangeeta}.

Shear variation for the Central Unzip (Figure 7d), clearly indicates melting 
of the central base pair occurs just after 1 ns. The base-pairs  next to the 
central pulled one (12th one) acquire large positive shear, while previous one 
(10th base pair) gets large negative shear and this feature continuous till the 
end. The base-pairs, which are away from the pulling point, maintain almost 
same value (near zero, cyan color) for much longer time. Similar structural 
changes can be observed in terms of stacking overlap value for all base 
positions as shown in Figure 7e. The overlap values of the 6th residue and 
14th residue appear to increase to value close to 35 {\r{A}$^{2}$} after 7 ns, 
indicating formation of secondary helix-loop-helix like structure within the 
separated single strands. This can be visualized by the snapshots (at 9 ns, 
12 ns and 18 ns) in lower panel of Figure 5. In terms of variation of twist 
value (Figure 7f) similar trend of separation of the strands starting from 
central pulling positions is observed. 

\section{Discussion }
In this study, we have tried to integrate structural transition of the 
system with different protocols of applied force. We found that, when 
pulling is applied at the axial direction (Axial Rupture) of the system, 
it experiences highest force compared to all other protocols. In this 
case highest force ($ \sim $  600 {kJ} {{mol$^{-1}$} {nm$^{-1}$}}) builds 
up to a point until certain critical interaction is broken. On the other 
hand, for the same direction pulling (Axial Stretch) the force continuously 
increases because the fixed end and pulled end are situated at the same 
strand. Nevertheless, ultimate disruption of most of the H-bonds in 
this case indicates that separation of the strands can also be achieved 
by this way of steering. In case of pulling perpendicular to the helix 
axis, the system unzips in usual way of strand separation. Here, opening 
of base-pairs are progressive and is achieved one by one from the pulling 
terminal. Due to this, during initial simulation time, force increases 
and then maintains almost same value over the pulling time. Among the 
protocols of pulling, Central Unzipping looks most different as compared 
to others. In this case, initially system experiences strong force, 
because of opening of first intact base-pairs and simultaneously unstacking 
the central base pair from both the sides of pulled nucleotides. But, 
eventually successive opening of progressive H-bonds requires much less 
force compare to all other cases of pulling. This is also revealed by the 
measurement of breaking of number of H-bonds. Hence this mode of opening 
can be viewed as most feasible as compared to all other possible protocols 
of siRNA strand separation.

We have also looked at the crystal structures of RNA double helical 
fragments bound to argonaute protein from Protein Data Bank \cite{pdb} to evaluate 
the most interactive structural signature of RNA. We found the protein bound 
double helical RNA strands
are of significant length (10 or more base-pairs) in 3HJF, 
3HK2, 3HM9, 4N47, 4NCB, 5AWH and 5UXO. We have analyzed hydrogen bonds between 
the protein and RNA using PyrHBfind software \cite{PyrHBfind} and have focused
on the strong ones.
The middle portion of the RNA duplex is found to be mostly interacting 
with protein residues by formation of very strong H-bonds involving 
negatively charged phosphate group of RNA and positively charged Lys or Arg 
residues of argonaute. In most of the cases, one of the ends of each 
strand is also found similarly anchored. e.g. 3HJF.pdb \cite{Wang} and 
3HK2.pdb \cite{Wang} have length of double helical region are 12 and 14 
respectively and around 3 to 4 bases of both the strand residing at the 
middle region of the helix in total forms H-bonds with the Protein. Hence, 
accepting the idea that more crowding will enable higher grip by protein 
and subsequently lead to rupture start-point, we can conclude that, central 
pulling must be most feasible phenomena in nature. However, crowding at one 
of the terminals may indicate that other mechanisms could also take place, 
though with lesser probability. The complete understanding of anchoring the 
siRNA at multiple points (center as well as the terminals) may require new methodological development to analyze the system. 
Again, the role of UU overhang cannot be 
detected in the unzipping of siRNA duplex, which might be involved in some 
other process. Among the helical parameters, we observed that for all the case 
of pulling, twist parameter is most sensitive during the opening of strands 
of siRNA duplex. Experimental determination of siRNA structure in Protein 
Data Bank is so far inadequate in number. In future perspective, it is 
necessary to study the effects different sequences and length for the 
system of siRNA and observe the changes associated. 

\section{acknowledgements}
We gratefully acknowledge SERB and DST, New Delhi India for their
financial supports through project numbers PDF/2015/000308 and PDF/2017/002110/CS. We also acknowledged CDAC for their computing 
support. Most of the simulations were performed in the cluster of 
CAPP-II project of DAE. We thank Prof. Sanjay Kumar of Banaras Hindu 
University and Prof. Rituparna Sinharoy of IISER-Kolkata for discussion 
and useful suggestions.

\end{document}